\documentclass[iop]{emulateapj}
\newcommand{\e}[1]{\ensuremath{\times 10^{#1}}}
\newcommand{\un}[1]{\ensuremath{\ \mathrm{#1}}}

\newcommand{\rsun}{\ensuremath{\ R_\odot}}
\newcommand{\rot}[1]{\ensuremath{\nabla\times {#1}}}
\newcommand{\diver}[1]{\ensuremath{\nabla\cdot {#1}}}
\slugcomment{To appear in ApJ}

\shorttitle{Coupling the solar dynamo and the corona}
\shortauthors{Pinto et al.}

\begin{document}

\title{Coupling the solar dynamo and the corona: \\
  wind properties, mass and momentum losses during an activity cycle}

\author{Rui F. Pinto and Allan Sacha Brun}
\affil{Laboratoire AIM Paris-Saclay, CEA/Irfu Universit\'e Paris-Diderot CNRS/ 
INSU, 91191 Gif-sur-Yvette, France
}
\email{rui.pinto@cea.fr}

\author{Laur\`ene Jouve}
\affil{
Universit\'e de Toulouse; UPS-OMP; IRAP; Toulouse, France \\
CNRS; IRAP; 14, avenue Edouard Belin, F-31400 Toulouse, France
}

\author{Roland Grappin}
\affil{
UMR 8102 CNRS-Observatoire de Paris -- Universit\'e Paris-Diderot,
Laboratoire Univers et Th\'eories, Observatoire de Meudon, 5 Place Jules Janssen, Bat 18, 92195 Meudon, France \\
Laboratoire de Physique des Plasmas,
Ecole Polytechnique, Palaiseau, France
}

 \begin{abstract}
   We study the connections between the sun's convection zone and the evolution of the solar wind and corona.
   We let  the magnetic fields generated by a 2.5D axisymmetric kinematic dynamo code (STELEM) evolve in a 2.5D axisymmetric coronal isothermal MHD code (DIP). The computations cover an 11 year activity cycle.
   The solar wind's asymptotic velocity varies in latitude and in time in good agreement with the available observations.
   The magnetic polarity reversal happens at different paces at different coronal heights.
   Overall sun's mass loss rate, momentum flux and magnetic braking torque vary considerably throughout the cycle.
   This cyclic modulation is determined by the latitudinal distribution of the sources of open flux and solar wind and the geometry of the Alfv\'en surface.
   Wind sources and braking torque application zones also vary accordingly.
\end{abstract}

\keywords{solar dynamo --- solar wind --- Sun: corona --- Sun:
  magnetic fields --- Sun: convective zone}

\maketitle

%%%%%%%%%%%%%%%%%%%%%%%%%%%%%%%%%%%%%%%%%%%%%%%%%%
% end of preamble %
%%%%%%%%%%%%%%%%%%%%%%%%%%%%%%%%%%%%%%%%%%%%%%%%%%

\section{Introduction}
\label{sec:intro}

The sun's magnetic field varies in time following a $22$-year cycle. The most visible manifestation of this periodic behaviour is the well-known $11$ years sunspot cycle, the sunspots themselves corresponding to high local concentrations (up to about $20\un{Mm}$ wide) of vertical magnetic field at the photosphere.
The sun's photosphere is in reality permeated by magnetic flux at several scales, and at all latitudes and longitudes.
This fact brings forward the idea that the magnetic field plays a role in coupling the dynamics of the inner layers of the sun and its atmosphere.
Studies about the convection zone (hereafter CZ) and atmospheric magnetic fields have mostly been done separately, nonetheless.
The first reason for this is that the photosphere essentially separates two regions of plasma with different regimes ($\beta\ll 1$ and $\beta\gg 1$) and disparate characteristic time and length scales, making any numerical investigations comprising them both a real challenge.
The second reason is that the actual coupling and/or cross-transport phenomena are still poorly understood (except perhaps at very small-scale, or in simple scenarios).
Finally, the third reason is linked to the use of different observational techniques for gathering data relating to the dense layers below the optically thick photosphere and to the essentially optically thin and rarefied atmosphere
\citep[see review by][]{zurbuchen_new_2007}.

The most remarkable exceptions are the numerical studies made in small cartesian domains which include the photospheric layers in the domain, the typical numerical domain size being of order of a few tens of$\un{Mm}$
(\citealt{
rempel_radiative_2009,
leenaarts_three-dimensional_2009,
martinez-sykora_twisted_2008,
voegler_simulations_2005}, among others; see also review by 
\citealt{wedemeyer-boehm_coupling_2009}).
These studies are unable, though, to capture the slowly varying magnetic field and solar wind's properties at a global scale.

The structure of the magnetic field inside the sun is unreachable by direct observation and/or measurement. Helioseismology techniques have been used, though, to deduce some of the properties of the magnetic field in the sun
\citep{kosovichev_properties_2006,kosovichev_active_2006,antia_suns_2000,antia_variation_2003}.
On the other hand, dynamo models are able to reproduce photospheric observables in greater detail.

Currently our understanding of the inner solar magnetism, its structure, evolution and origin rely on multi-D models of solar dynamo.
This ``fluid'' dynamo is at the origin of the intense magnetic activity of the sun.
In particular, the solar interface dynamo paradigm has received much attention \citep[][]{parker_solar_1993}.
This model assumes that the locations of generation of the toroidal and poloidal global field are separated, with the tachocline playing a central role in organising toroidal field.

The classical explanation for the cyclic activity of the large scale 
magnetic field is that a dynamo process acts in the solar interior to 
regenerate the three components of the magnetic field and sustain them 
against ohmic dissipation. The inductive action of the complex fluid 
motions would thus be responsible for the vigorous regeneration of 
magnetic fields and for its nonlinear evolution in the solar interior 
(see \citealt{charbonneau_dynamo_2010} and \citealt{miesch_large-scale_2005} for recent reviews on the 
subject).
Understanding how these complex physical processes operating 
in the solar turbulent plasma non-linearly interact is very challenging. 
One successful and powerful approach is to rely on multi-dimensional 
magnetohydrodynamic (MHD) simulations. 
In this context, two types of 
numerical experiments have been performed since the 70's: kinematic 
mean-field axisymmetric dynamo models which solve only the mean 
induction equation 
\citep{steenbeck_dynamo_1969,roberts_kinematic_1972,stix_differential_1976,krause_mean-field_1980}
 and full 3D global models which explicitly solve the full set of MHD equations 
\citep{gilman_dynamically_1983,glatzmaier_numerical_1985,cattaneo_origin_1999,brun_global-scale_2004}.

Clearly, both approaches are complementary and are needed to better understand the magnetic solar activity.
Recent progress have been made with 3D numerical models of magnetic stars.
Large-scale magnetic cycles are starting to be found in simulations of the Sun, such as those performed by \citet{ghizaru_magnetic_2010}, or of solar-like stars \citep{brown_submitted_2011}.
However, butterfly diagrams similar to the solar observations --- with the poloidal field field reversal happening when the toroidal field is at its maximum --- are still difficult to reproduce.
Kinematic mean-field models and their associated simplifying assumptions 
have thus been used extensively to reproduce several features of the 
large-scale solar cycle. 
In particular the use of a differential rotation profile inferred from helioseismology associated with an alpha-effect (due to the helical turbulence of the stellar convective envelope) antisymmetric with respect to the equator enabled \citet{charbonneau_solar_1997} to produce a solar-like butterfly diagram with the ingredients of the modern interface dynamo.
A similar model will be considered in this work to catch the large-scale behaviour of the inner solar magnetic field.

Above the photosphere, a very complex structure comprising open magnetic flux-tubes (coronal holes) and magnetic loops with different length-scales arises (magnetic carpet, canopy, coronal loops)
\citep[][and references therein]{stix_sun:_2002,aschwanden_physics_2005}.
The wealth of ground and space based observational data gathered in the latest years has been bringing up a great insight on the magnetic structure in the lower atmospheric layers (chromosphere and lower corona).
The properties of the magnetic field in the (optically thin) higher corona are harder to determine observationally, though.
Space-borne \emph{in-situ} measurements may introduce further constraints about how the solar wind properties connect to the coronal structures.
Remarkable examples of these were given by the Ulysses consecutive polar orbits 
\citep{mccomas_three-dimensional_2003,
  issautier_solar_2004,
  issautier_electron_2008}.
The SoHO spacecraft further provided complementary \emph{in-situ} and solar surface (and low coronal) observations.
Future missions such as the Solar Orbiter should provide more refined data to complete the scenario.
These should combining high resolution imagery and spectroscopic data, magnetogram and \emph{in-situ} measurements of the wind's properties in an orbit with very close helio-synchronous passes and varying orbital inclinations.

%%%%%%
The solar wind's outflow properties depend on the particular geometry of the coronal field at each moment of the solar cycle.
The amplitude and latitudinal distribution of the solar wind velocity and mass flux depend on parameters such as the positions of the wind sources at the surface, the local magnetic field strength and the expansion factor of each particular coronal flux-tube the wind flows along.
The details about the coronal heating mechanisms, most notably the amplitude and location of energy dissipation, are also of importance.
\citet{leer_energy_1980} pointed out that heating below or above the critical sonic point may produce different effects on the wind velocity, and that high speed winds require energy deposition in the supersonic region.
\citet{hansteen_coronal_1995} showed that the mass flux does not depend strongly on the mode and location of energy deposition, but rather on the amplitude of the energy flux.
Note also that the heat flux profile could be related to the magnetic field's amplitude and geometry \citep{cranmer_self-consistent_2007,pinto_time-dependent_2009}.
The resulting angular momentum losses and wind's braking torque (applied on the sun) depend on wind's velocity and mass flux together with the (time varying) geometry of the Alfv\'en surface (the surface at which the wind's velocity equals the Alfv\'en velocity; see definitions in \S\ref{sec:massflux}).
Some authors have studied the influence of the topology of coronal field in the properties of stellar winds and resulting braking torques, but mostly by using simplified configurations (e.g, dipolar versus quadrupolar field, as in \citealt{matt_accretion-powered_2008-1}).
Others have furthermore considered how the presence of strong magnetic spots (localised magnetic flux enhancements) specifically affects the angular momentum loss rate \citep[e.g,][]{aibeo_magnetic_2007,cohen_effect_2009}.

More detailed studies of the global solar magnetic structure usually adopt
surface magnetogram data for the radial component of the field as a lower boundary.
The atmospheric magnetic field's geometry is deduced using potential field extrapolation techniques. 
The coronal and heliospheric hydromagnetic conditions are then deduced either by using a set of semi-empirical relations
\citep{arge_improvement_2000,luhmann_solar_2002,wang_potential_1992,schrijver_photospheric_2003},
 or by finding stable MHD solutions for the mapped fields
\citep[e.g,][]{usmanov_global_1993,hu_three-dimensional_2008}.
The usage of magnetogram data as lower boundary conditions encounters a few difficulties.
Namely, only the line-of-sight component of the field is available, making the estimation of the radial components difficult at the polar regions. 
The standard PFSS technique (standing for ``potential field source surface'') further assumes the surface field is strictly radial.
One still needs to assume an outer boundary condition which imposes the magnetic field to be purely radial as well at a \emph{source-surface} (typically, a spherical surface placed at about $r=2.5\un{\rsun}$).
This condition emulates the field-line opening usually caused by the solar wind flow, while keeping the global field at its lowest (current-free) energy state.
These studies essentially provide quasi-stationary predictions of the coronal fields at a given moment.
Despite all the simplifications and assumptions this family of models has been successful in predicting qualitatively the coronal magnetic field's topology (observed, for example, in white light during solar eclipses) and became a standard to which other types of model can be compared.
Their MHD counterparts are limited in different ways.
Beyond the much increased computational power they require, they rely on phenomenological assumptions for the heat transport in the corona (as the heating processes are still under debate).
These models show, nevertheless, some quantitative differences when compared to the PFSS ones.
The relative sizes of the streamers can be different (usually more elongated) and the heliospheric magnetic field tends to be more uniform (as it is known to be).
See, for example, \citet{riley_comparison_2006} for a comparison between these two families of models.
A third type of models --- nonlinear force-free field (NLFFF) models --- places itself somehow in between the two previously described, in the sense that it tries to add some complexity to the PFSS scenario while avoiding the issues found by global MHD computations.
These require, though, good quality vector magnetogram data at the surface of the sun and still find some difficulty in accurately predicting real solar features \citep{derosa_critical_2009}.

In this paper, we investigate the influence of the cyclic evolution of the large-scale magnetic field produced by the solar dynamo on the solar wind properties.
We chose not to perform this study based on surface magnetogram magnetic field data
\citep[as in][for example]{
  wang_topological_2003,
  hu_three-dimensional_2008,
  luhmann_solar_2009}, 
but rather pursue a different path: our source fields are the result of well-tested and validated kinematic dynamo models \citep{jouve_solar_2008}.
An isothermal MHD model of the solar wind and corona is used to produce a temporal sequence of steady-states spanning a complete activity cycle.
The focus here is on estimating the wind's velocity, mass and angular momentum flux spatial profiles as they vary during an activity cycle in response to the variations of the magnetic field's topology, rather than reproducing a particular solar cycle.
The variability of global properties such as the sun's mass loss rate and wind's breaking torque are studied in regards to the activity cycle.
Also, we expect to gain a deeper insight on the connections between the sub-surface and coronal physical processes by proceeding this way.

The remainder of this manuscript is organised as follows: the methods
and numerical codes used are presented in \S\ref{sec:method}. The
results are described thoroughly in \S\ref{sec:magfield} (coronal magnetic field evolution), \S\ref{sec:windspeed} (solar wind speed) and \S\ref{sec:massflux} (mass and angular momentum flux, breaking torque).
A discussion follows in \S\ref{sec:disc} and a brief summary in \S\ref{sec:summary}.

\section{Coupling solar dynamo and wind models}
\label{sec:method}

We used two 2.5D axisymmetric MHD codes.
The first one ---  STELEM \citep[][and \S\ref{sec:stelem}]{jouve_role_2007} --- computes mean-field kinematic MHD dynamo solutions given the meridional circulation and differential rotation profiles in the solar convection zone.
A Babcock-Leighton or an $\alpha$ source terms generate the poloidal magnetic field.
The second code --- DIP \citep[][and \S\ref{sec:dip}]{grappin_alfven_2000} --- computes the temporal evolution of an MHD solar corona with a self-consistent wind.
In the latter, the magnetic field separates into two components: an imposed stationary \emph{external} potential field $\mathbf{B}^0$ (whose sources lie within the sun), and an induced component produced by the flows within the numerical domain.
The dynamo magnetic fields produced by STELEM match a potential field at the surface of the sun.
This potential source field can then be directly transmitted to DIP as the \emph{external} component of the coronal magnetic field.
Further details about both numerical codes are given hereafter, in \S\ref{sec:stelem} and \S\ref{sec:dip}. Details about the coupling of the two codes are given in \S\ref{sec:couplingmethod}.

%%% fig 1
\begin{figure}[!h]
  \centering
  \includegraphics[width=.4\linewidth]{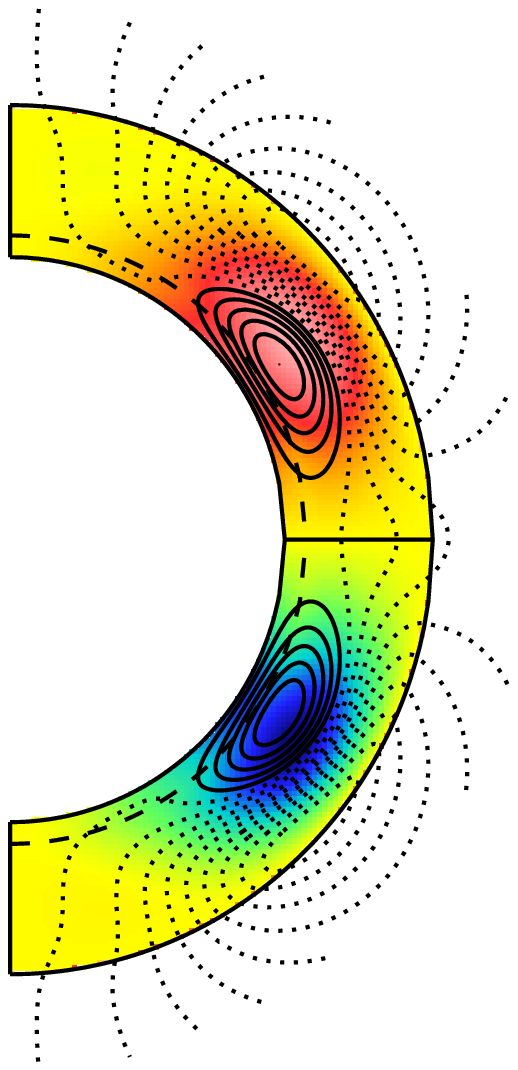}
  \includegraphics[width=.4\linewidth,clip=true,trim=0 0 0 15]{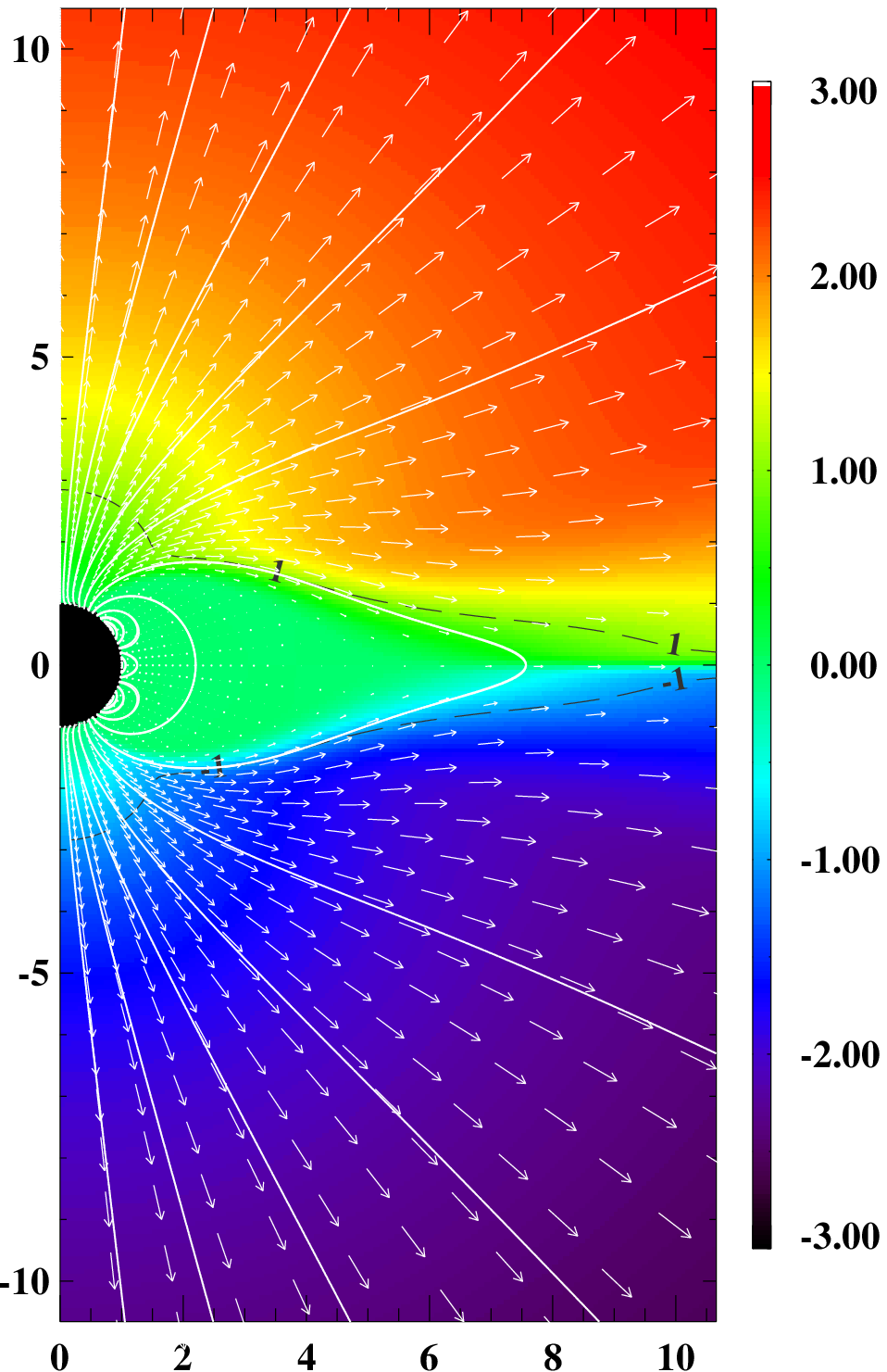}
  \caption{Representation of the numerical domains of STELEM (left) and DIP (right) at the same instant.
    The colour scale in the left figure shows the toroidal magnetic field $B_\phi$ in the convection zone. Black lines are (poloidal) magnetic field lines (dotted and continuous lines indicates CW and CCW field-lines, respectively). 
    The white lines in the right figure are magnetic field lines, while the colour-scale represents both the wind's velocity in units of Mach number and the open field's polarity (blue is negative, red is positive).}
  \label{fig:dominios}
\end{figure}

\subsection{STELEM: a representative dynamo solution}
\label{sec:stelem}

To investigate the global solar cycle features produced by dynamo models, we start from the hydromagnetic induction equation,  
governing the evolution of the magnetic field ${\bf B}$ in  
response to advection by a flow field ${\bf U}$ and resistive  
dissipation.

$$
\frac{\partial {\bf B}}{\partial t}=\nabla\times ({\bf U} \times{\bf  
B})-\nabla\times(\eta\nabla\times{\bf B})
$$

As we are working in the framework of mean-field theory, we express  
both magnetic and velocity fields as a sum of large-scale (that will  
correspond to the mean field) and small-scale (associated with fluid  
turbulence) contributions. Averaging over some suitably chosen  
intermediate scale makes it possible to write two distinct induction  
equations for the mean and the fluctuating parts of the magnetic  
field. 

A closure relation is then used to express the mean electromotive force in terms of the mean magnetic field, leading to a simplified mean-field equation \citep{moffatt_magnetic_1978}.
The prescribed mean velocity field will here only consist in its longitudinal component (the solar differential rotation) and the magnetic diffusivity is assumed to be constant.
 The source term for the poloidal field is linked to the turbulent helical motions within the convection zone.
We thus obtain a simple $\alpha\Omega$ dynamo model with constant diffusivity.

Working in spherical coordinates and under the assumption of  
axisymmetry, we write the total mean magnetic field {\bf B} as
$$
{{\bf B}}(r,\theta,t)=\nabla\times (A_{\phi}(r,\theta,t) \hat {\bf  
e}_{\phi})+B_{\phi}(r,\theta,t) \hat {\bf e}_{\phi}
$$

Reintroducing this poloidal/toroidal decomposition of the field in the  
mean-field induction equation, we get two coupled partial differential  
equations, one involving the poloidal potential $A_{\phi}$ and the  
other concerning the toroidal field $B_{\phi}$

\begin{equation}
  \label{eqA2}
  \frac{\partial {A_{\phi}}}{\partial t}= C_{\alpha}\alpha B_{\phi} 
  + (\nabla^{2}-\frac{1}{\varpi^{2}})A_{\phi}
\end{equation}

\begin{eqnarray}
  \label{eqB2}
  \frac{\partial {B_{\phi}}}{\partial t}&=& C_{\Omega}\varpi(\nabla 
  \times(\varpi A_{\phi}{\bf \hat{e}}_{\phi}))\cdot\nabla\Omega \nonumber \\ 
  &+& (\nabla^{2}-\frac{1}{\varpi^{2}})B_{\phi}
\end{eqnarray}
where $\varpi=r\sin\theta$,
$\Omega$ is the differential rotation and $\alpha$ is the $\alpha$-effect.

All these quantities are now dimensionless, thanks to the presence  
of the two Reynolds numbers $C_{\Omega}=\Omega_{eq}R_{\odot}^2/\eta_{t}$,  
and $C_{\alpha}=\alpha_{0}R_{\odot}/\eta_{t}$,
$\eta_t$ being the turbulent magnetic diffusivity, $\Omega_{eq}$ a measure of the rotation rate at the equator and $\alpha_0$ a measure of the intensity of the alpha-effect.
The product of these two numbers gives us the dynamo number which measures the efficiency of the source terms to make the magnetic energy grow in time against Ohmic diffusion.

Equations (\ref{eqA2}) and (\ref{eqB2}) are solved in an annular  
meridional cut with the colatitude $\theta$ $\in [0,\pi]$ and the  
radius $r \in [0.65,1]\un{R_{\odot}}$ i.e from slightly below the  
tachocline (e.g. $r=0.7\un{R_{\odot}}$) up to the solar surface (see Figure \ref{fig:dominios}, left panel).
At $\theta=0$ and $\theta=\pi$ boundaries, both $A_{\phi}$ and $B_{\phi}$  
are set to 0. At $r=0.65\un{R_{\odot}}$, we compute a perfect conductor  
condition.
At the upper boundary,  we smoothly match our solution to an external  
potential field, i.e. we have vacuum for $r \geq R_{\odot}$.

The STELEM (STellar ELEMents) code uses a finite element method in  
space and a
third order scheme in time 
(\citealt{burnett_finite_1987}, \citealt{jouve_role_2007}).

The principle of the finite element method is to look for solutions of the weak formulation of the equations, here equations (\ref{eqA2}) and (\ref{eqB2}). Those solutions are taken to be linear combinations of well-chosen trial functions. In our case, they are Lagrange polynomials of degree 1 and the elements are rectangles in the $(-\cos\theta,r)$ plane. Applying this spatial method to our equations results in a system of first order \emph{ODE}s in time governing the evolution of the coefficients of the linear combinations.
The temporal scheme that we use is adapted from \citet{spalart_spectral_1991}. 
It is similar to a Runge-Kutta 3 method and is thus explicit and of order 3.

The STELEM code is here used to produce a cyclic dynamo field within the model convection zone, whose values at the top of our domain will be reintroduced in the DIP code described below. 
We chose a simple $\alpha\Omega$ dynamo model to produce our representative solution, with physical source terms consistent with observations.
The rotation profile is deduced from helioseismic inversions and the $\alpha$-effect is antisymmetric with respect to the equator and positive in the Northern hemisphere (in agreement with the preferred handedness of the turbulent helical convective motions).
This is similar to case A of \citep{jouve_solar_2008}.
The following values were used for the various parameters: $C_{\alpha}=0.385$ (critical value for which dynamo action just starts to compete against Ohmic diffusion), $C_{\Omega}=1.4 \times 10^5$ (corresponding to a value of $\Omega_{eq}=460 \, \rm nHz$, in agreement with observations). The spatial resolution is $128\times 128$.  

%%% fig 2
\begin{figure}[h!]
  \centering
  \includegraphics[width=\linewidth]{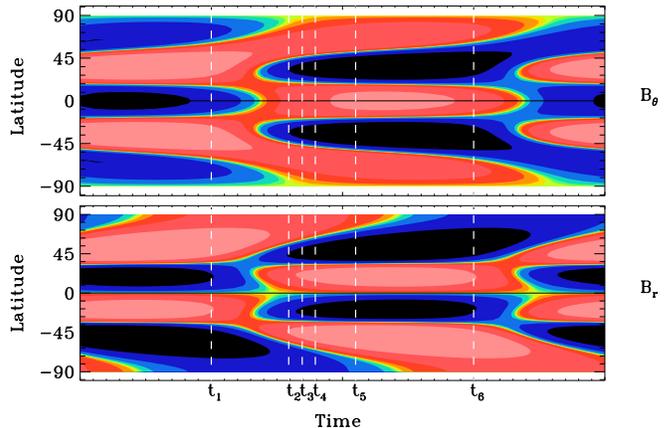}
  \caption{Time-latitude diagram of the surface field.
    The upper panel shows $B_\theta$ while the lower panel shows $B_r$. 
    Red (blue) colours indicate positive (negative) values of the field. 
    The white dashed lines span the time interval which was chosen to be incorporated in the DIP code to compute the solar wind evolution and mark the instants shown in the following figures.
  }
  \label{fig:papillon}
\end{figure}

Figure \ref{fig:papillon} shows the temporal evolution of both the latitudinal and the radial field at the surface, at all latitudes.
We have here the evidence of a cyclic magnetic field: the surface poloidal field changes sign regularly and at different times depending on the latitude of interest.
This simple model does not aim at reproducing the solar dynamo in its finest details.
The goal here is rather to assess the influence of a cyclic poloidal background field (having the same general properties as that in the sun) on the wind's properties.
The dashed white lines labelled $t_1$ and $t_6$ in the figure indicate the beginning and the end of the particular time interval used to compute the solar wind evolution.
This time interval spans an entire cycle period and is roughly located between two activity minima.

\subsection{DIP: wind model}
\label{sec:dip}
The DIP code is a 2.5 D axisymmetric model of the solar
corona obeying the compressible MHD equations for a one-fluid, isothermal and fully ionised plasma.
The continuity and momentum equations are 
\begin{eqnarray}
  \label{eq:mhd1}
  \partial_t \rho & + & \diver{\rho\mathbf{u}} =0  \\
  \partial_t \mathbf{u} & + & \left(\mathbf{u}\cdot\nabla\right)
  \mathbf{u} = \nonumber \\ 
  & & -\frac{\nabla P}{\rho} +
  \frac{\mathbf{J}\times\mathbf{B}}{\mu_0\rho} -
  \mathbf{g} + \nu\nabla^2\mathbf{u}\ . \nonumber 
\end{eqnarray}
We set $\gamma=1$ and a uniform temperature $T = 1.3\un{MK}$ in all the corona.
The gas pressure is deduced from the equation of state 
\[  P = \frac{2}{m_H}\rho k_BT\ , \]
valid for a fully ionised hydrogen plasma.
The isothermal approximation discards a complete treatment of the energy fluxes in the corona.
But we should note firstly that the corona is nearly isothermal (at least in the first $12\un{\rsun}$). 
Secondly, assuming $\gamma=1$ is a proxy to the actual thermal state of the coronal plasma, summing-up the combined effects of the thermal conduction and the (still debated) heating source.
The choice of a particular value for the coronal temperature is somewhat arbitrary, though.
The temperature $T = 1.3\un{MK}$ chosen here is justified both by representing an average coronal value in the first $\sim 10 \un{\rsun}$ \citep[see, e.g, results by][among others]%
{hansteen_coronal_1995,endeve_two-dimensional_2003,guhathakurta_semiempirically_2006,pinto_time-dependent_2009} and empirically as it produces correct wind solutions.
As discussed later in the text, the wind's mass flux depends both on the coronal temperature and on wind geometrical expansion factors.
We focus here on the geometrical effects linked to the activity cycle and keep our reference temperature fixed throughout the simulations.

The magnetic field $\mathbf{B}$ decomposes into a potential
\emph{external} component $\mathbf{B^0}$ and on a component $\mathbf{b}$ induced by the flows.
The former only has poloidal components ($r, \theta$), while the latter comprises both poloidal and toroidal components.
The total magnetic field is then defined as
\begin{equation}
  \label{eq:btotal}
  \mathbf{B} = \mathbf{B}^0 +  \mathbf{b}_{pol} + \mathbf{b}_\phi \ .
\end{equation}
Furthermore, $\mathbf{b}_{pol} = \rot{\mathbf{\Phi}}$.
We integrate the evolution equation for the magnetic potential $\mathbf{\Phi}$
\begin{equation}
  \label{eq:phi}
  \partial_t \mathbf{\Phi} = 
  u_r B_\theta - B_r u_\theta + \eta \nabla^2\mathbf{\Phi}\ .
\end{equation}
The poloidal components of $\mathbf{b}$ are then computed as
\begin{eqnarray*}
  b_r &=& \frac{1}{r}\left( \Phi \cot\theta + \partial_\theta
    \Phi \right) \ , \\
  b_\theta &=&  - \frac{\Phi}{r} -  \partial_r \Phi \ .
\end{eqnarray*}
The azimuthal component of the induced field comes directly from the induction equation (in the $\phi$ direction only), that is
\[  \partial_t b_\phi = \left[\rot{\left(\mathbf{u}\times\mathbf{B}\right)} +
  \eta\nabla^2\mathbf{B}\right]_\phi \ . \]
This method guarantees that the solenoidal condition $\diver{\mathbf{B}}=0$ is satisfied at all times.
This papers focus only on the poloidal components of the coronal field, though (as explained hereafter, in \S\ref{sec:couplingmethod}).

The diffusive terms are adapted so that grid scale ($\Delta l$)
fluctuations are correctly damped.
The kinematic viscosity is defined as $\nu=\nu_0\left(\Delta l/\Delta l_0 \right)^2$, typically with $\nu_0=2\times 10^{14}\un{cm^2\cdot s^{-1}}$ and
$0.01\lesssim\left(\Delta l/\Delta l_0 \right)^2\lesssim10$. 
The simulations presented in this paper were performed with a vanishing $\eta$, so as to approach as much as possible the limit of ideal non-resistive MHD.
Additional diffusive terms are used.
These are implicit numerical filters (applied over $\mathbf{u}$, $\mathbf{b}$ and $\Phi$) which dissipate mostly at the grid scale and minimise the dissipation of large scales fluctuations \citep{lele_compact_1992}.
This filtering scheme allows the diffusive parameters $\nu$ and $\eta$ to be lowered while avoiding spurious Gibbs fluctuations, and allow for lower mid-scale damping \citep{grappin_turbulent_2001}.
Note that actual kinetic dissipation should happen at scales much smaller than the grid size, anyway.

We used a $512^2$ grid which is uniform in latitude and non-uniform in radius.
The grid's cells radial extent is $\delta r=6.5\e{-3}\un{\rsun}$ at the lower boundary and $\delta r=1.0\e{-1}\un{\rsun}$ at the upper boundary.

Both the upper and the lower numerical boundaries (respectively at
$r=15$ and $1.01\rsun$) are transparent, \emph{i.e}, mass can flow through them and waves are not spuriously reflected there (see Figure \ref{fig:dominios}, right panel).
This is achieved by writing the MHD equations in their characteristic form.
In simple terms, this consists in projecting the system of equations in terms of the
\emph{primitive} quantities $\{ \rho,\mathbf{U},\mathbf{B}\}$ into an
equivalent system defined in terms of characteristic
variables $L_i$
\citep{thompson_time_1987, vanajakshi_boundary_1989, poinsot_boundary_1992,roe_notes_1996}.
The evolution equation for these variables explicitly
define the time-evolution of the MHD system in terms of wave-modes
propagating upwards and downwards, or equivalently, incoming and
outgoing modes. We must then constraint all the incoming modes, and
let the others free.
Plasma is free to flow through and its properties to vary in time, the actual boundary values at each moment depending on the current state of the system.
The mass fluxes and velocities (or their gradients) are, therefore, not arbitrarily set at the numerical boundaries.
This is a critical feature: the actual solar wind velocities and mass flux at each point of the surface will automatically match each unique transonic and transalfv\'enic wind solution.
There is no need to set in advance arbitrary values for the density, velocity or mass flux at the numerical boundaries.
This means no spurious boundary layers will form there and that the resulting mass fluxes will not be artificially constrained.
We note that the resulting surface density and mass flux do not vary by large factors (not more than $\sim 10\%$ in all latitudinal domain, and less than $6\%$ within open field regions at all moments of the cycle).
Furthermore, this type of boundary conditions better describes dynamic finite-frequency phenomena than any type of non-transparent (i.e, ``rigid'') conditions \citep{grappin_mhd_2008}.

The solar wind develops into a stable transonic and transalf\'enic solution
in the open field regions after a pressure (sonic) perturbation is applied at the outer boundary.
This perturbation propagates inwards until both the sonic and alfv\'enic surfaces appear (and are enclosed in) within the domain.
From then on, the solar wind profiles evolve towards a unique stable state \citep{velli_supersonic_1994,grappin_thermal_1997}.
Note that the frontiers between open (coronal holes) and closed flux regions (streamers) are not set in advance, but they are rather a result of the competition between the magnetic tension and the dynamical pressure in the wind flow.
In other words, the solar wind settles where the magnetic field is incapable of containing the outward flow and forces the magnetic field to align with the radial direction.

Other aspects of the numerical model are more thoroughly discussed in
\citet{grappin_alfven_2000}.

\subsection{Coupling method}
\label{sec:couplingmethod}

We used a time-series $\lbrace B_r^0\left(t\right), B_\theta^0\left(t\right)\rbrace$ derived from a STELEM run \citep{jouve_role_2007} which correctly produces a cyclic behaviour of the poloidal magnetic field.
We used a simple $\alpha\Omega$ dynamo model with source terms for both the toroidal field and the potential $A_\phi$.

The time-series was scaled in order to span an $11$-year period, sampled with a time step of $6$ months.
The vertical dashed white lines labelled $t_1$ and $t_6$ in Figure \ref{fig:papillon} show the beginning and end of the time-series ($t_1=0$ and $t_6=11\un{yr}$).
These are the moments when the poloidal field is at its simplest configuration (i.e, lowest multipolar order).
For simplicity, we will call these moments ``activity minima'', while the magnetic field's polarity inversion phase (between lines $t_2$ and $t_5$ in the same figure) will be called ``activity maximum'' hereafter.
The time-series comprises $22$ equally spaced samples, but we will refer to a subset of $6$ samples ($t_1$ to $t_6$) for illustrative purposes throughout the text.
We let the solar wind fully develop in the domain and reach a steady state (see \S\ref{sec:dip} for further details on the formation of the wind.)
The amplitude of the external magnetic field $\mathbf{B}^0$ was scaled so that the total field would match coronal amplitudes (the scaling factor being kept constant during the whole cycle).
Then, for each consecutive sample, we substituted directly $\mathbf{B}^0_i \longmapsto \mathbf{B}^0_{i+1}$.
The system was again let free to relax and attain a new steady state for each iteration $i$.
The stability of the relaxation method was tested so that we could be sure that each relaxed steady-state solution did not depend on the history.
That is, we found no hysteresis when cycling back and forth through different states (given that we let the system relax at each stage).
Different permutations of the original time-series $i=0,1,\ldots,N$ also lead to the exact same steady-states as long as each solution's set of parameters was kept the same.
The wind solution for each instant of time $i$ depends only on the corresponding $\mathbf{B}^0_i$.

Some of the runs were performed at different grid resolutions, and numerical convergence was verified.

Both numerical codes are time-dependent in nature, but the result of the whole coupling procedure is a sequence of steady-state solutions.
Our procedure generates a map of the poloidal coronal magnetic field and wind flow during an activity cycle rather than the dynamical evolution of a particular event.

Toroidal fields and flows will not be considered in this paper.
Concerning rotation, the sun is a slow magnetic rotator, and therefore the magneto-rotational effects on the poloidal wind flow are negligible \citep{belcher_magnetic_1976}.

\section{Coronal magnetic field}
\label{sec:magfield}

%%% fig 3
\begin{figure*}[!ht]
  \centering
  \includegraphics[width=.9\linewidth]{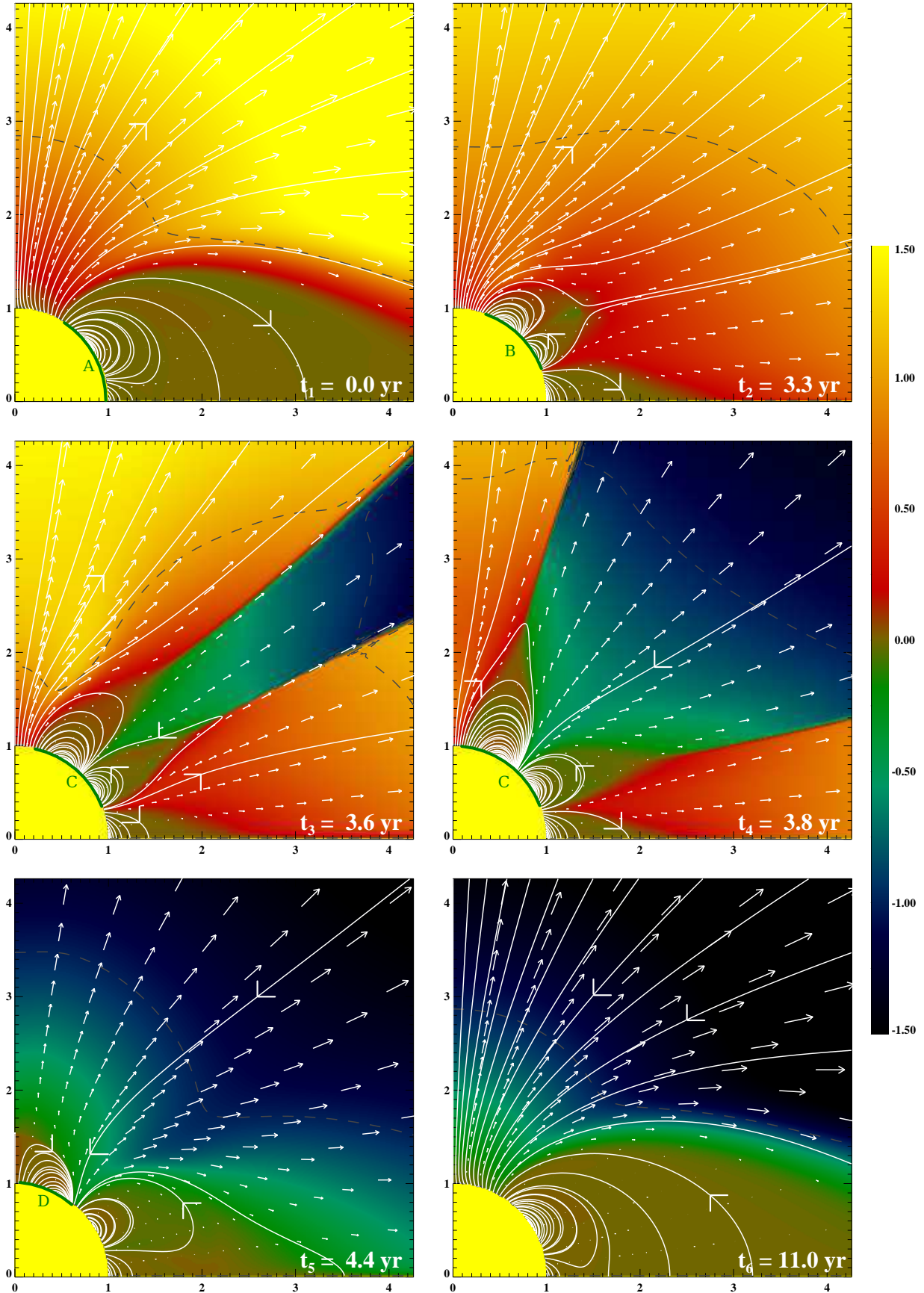}
    \caption{Snapshots of the evolution of the corona during the solar
      cycle (only the first $\sim 4\rsun$ and northern hemisphere are
      shown) at the instants $t_1$ to $t_6$ shown in Figure \ref{fig:papillon}.
      White lines are magnetic field lines.
      The colorscale represents the quantity 
      $\frac{\mathbf{u}\cdot\mathbf{B}}{c_s \|\mathbf{B}\|}$, that is, the  
      wind flow velocity projected onto the signed magnetic field in units
      of Mach number.
      This quantity traces the $B$-field's
      polarity in the open field regions. Red/orange means positive
      polarity, while green/blue means negative polarity. The large
      arrowheads show the local $\mathbf{B}$-field orientation. The
      grey contour shows the sonic surface.
      The letters A, B, C and D indicate the positions of particular magnetic structures which we refer to in the text.
    }
  \label{fig:field}
\end{figure*}

%%% fig 4
\begin{figure}%[]
  \centering
  \includegraphics[width=0.95\linewidth]{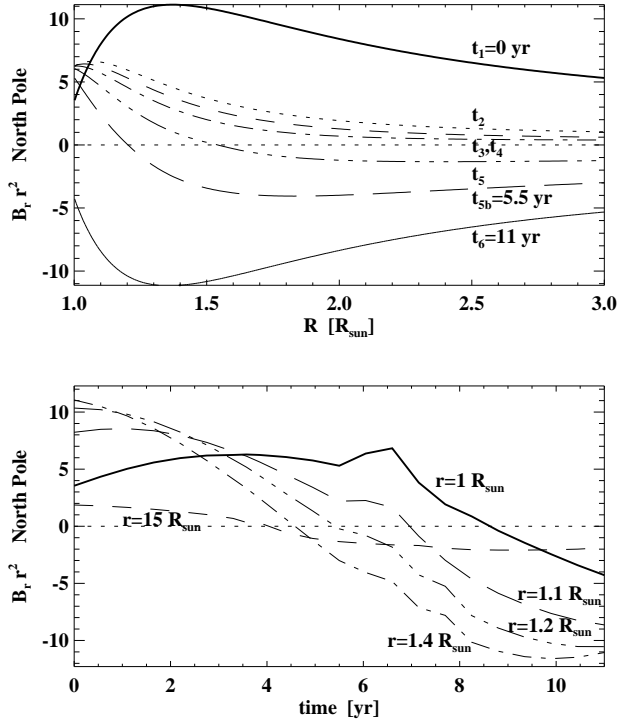}
  \caption{Top panel: Polar magnetic field as a function of radius at different moments of the cycle.
    The instants represented are the same as in the panels in Figure \ref{fig:field} ($t_1$ to $t_6$) plus an additional $t_{5b}=5.5\un{yr}$ for completeness.
    Bottom panel: $B_r r^2$ at the north pole as a function of time.
    $B$ is in units of $\mathrm{G}$ and $r$ is in units of $\mathrm{\rsun}$.
    Each curve corresponds to a different height.
    %($r = 1,\ 1.1,\ 1.2,\ 1.4,\ 15\un{\rsun}$).
    The $r^2$ factor accounts for the field's decay due to a purely radial expansion (note that $B_r r^2$ decays faster in the lower part of the domain, but not above).
    The polarity inversion at the surface is delayed with respect to higher coronal heights.
    This is due to the slowly progressing closing down of polar fields, as seen in Figure \ref{fig:field}.
  }
  \label{fig:brpole}
\end{figure}

%%% fig 5
\begin{figure}%[]
  \centering
  \includegraphics[width=0.95\linewidth]{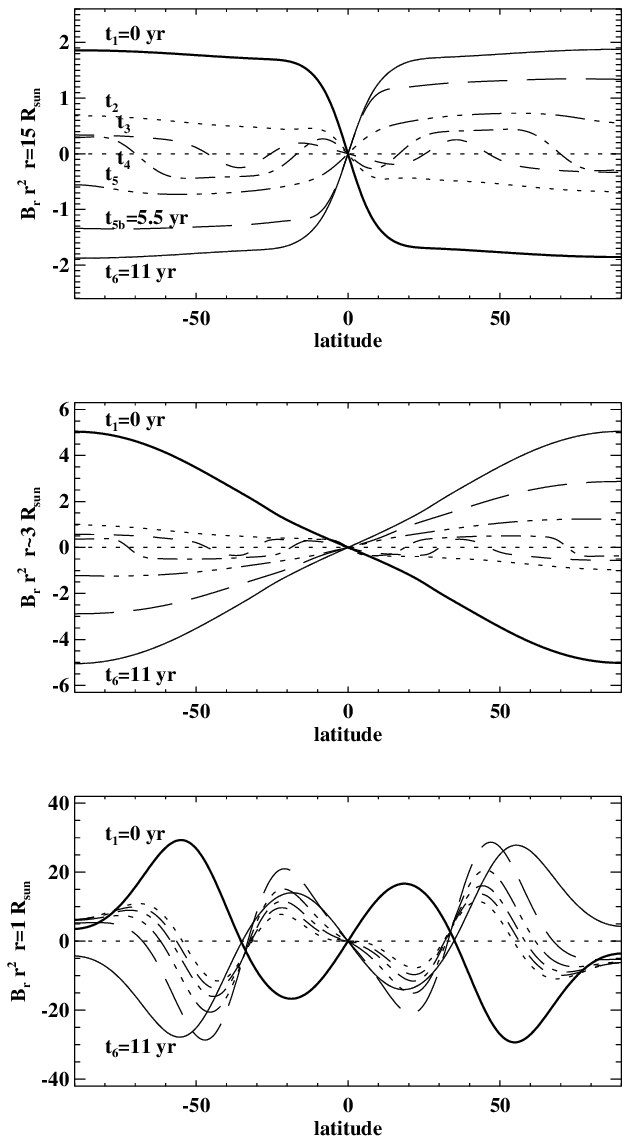}
  \caption{$B_r r^2$ as a function of latitude at different heights, and at different moments of the cycle.
    The $r^2$ factor accounts for the field's decay due to a purely radial expansion, as in the previous figure.
    $B$ is in units of $\mathrm{G}$ and $r$ is in units of $\mathrm{\rsun}$.
    The instants represented are the same as in the panels in Figure \ref{fig:field} ($t_1$ to $t_6$) plus an additional $t_{5b}=5.5\un{yr}$ for completeness.
    Faraway from the sun, the radial magnetic field is mostly uniform in latitude (except around current sheets, where it changes sign), independently of the complexity of the surface field.
  }
  \label{fig:brlatitude}
\end{figure}

The temporal evolution of the corona and solar wind in response to the dynamo field variations is shown in Figure \ref{fig:field}.
Only the first $4\un{\rsun}$ of the northern hemisphere are displayed at six different instants of the activity cycle, corresponding to the $t_1,\ t_2,\ \ldots,\ t_6$ lines in Figure \ref{fig:papillon}. That is: 
$t=0,\ 3.3,\ 3.6,\ 3.8,\ 4.4,\ 11$ years.
The colour-scale represents the solar wind's velocity projected on the unit magnetic field vector in units of Mach number, that is ${\mathbf{v}\cdot\mathbf{B}}/{\left(c_s \|\mathbf{B}\|\right)}$. Orange/yellow and blue/green shades therefore trace different $\mathbf{B}$-field polarities in the open field regions (respectively, $\mathbf{u}\cdot\mathbf{B}$ positive and negative).
The sharp transitions between positive and negative polarities in this figure outline current sheets (note that the wind flow does not change sign across these transitions, but $B_r$ does).

Some elements and characteristics are observed consistently throughout the whole cycle.
Higher concentrations of magnetic flux at the surface (or equivalently, of current below the surface) translate into coronal loop arcade systems.
Strong flux concentrations appearing in coronal holes shape up as 
\emph{helmet streamers}, and end in a current sheet which extends outwards.
Smaller flux concentrations embedded in unipolar flux regions form nearly symmetric bipolar structures with no current sheet
(as the ``giant polar plumes'' in \citealt{pinto_coronal_2010} and the ``pseudo-streamers'' in \citealt{wang_solar_2007}).
Loop structures placed inside stagnant zones (\emph{i.e} with no wind flow, as inside large streamers) mostly maintain their potential field configuration.

We start the computation at the moment when the magnetic structure of the corona is at its simplest (Figure~\ref{fig:field}, first panel, $t_1$; also first white line in Figure \ref{fig:papillon}). 
One and only large equatorial streamer extends from the surface up to nearly $5\un{\rsun}$, where the heliospheric current sheet starts.
The frontiers between the streamer and the coronal holes cross the surface at latitudes $-60^\circ$ and $+60^\circ$.
The letter A indicates the latitudinal extent of the streamer.
The open magnetic field has positive polarity in the northern coronal hole and negative polarity in the southern coronal hole. The streamer itself is divided into four magnetic connectivity regions around one X-type null point.
The null point itself is located over the equator at $r=1.4\un{\rsun}$.
The four connectivity regions mentioned above are then the group of small equatorial loop arcades place below the null point, the group of larger equatorial arcades above the null and filling up most of the streamer, and the two groups of arcades to the north and south of the null (note that Figure \ref{fig:field} only shows one hemisphere, and that the system is symmetric with respect to the equator).
As the solar cycle starts moving away from the minimum, new flux concentrations emerge at mid-latitudes (showing up as new groups of coronal loops inside the equatorial streamer). 
These new structures slowly migrate polewards, attaining the streamer boundaries at about $t=2.5\un{yr}$. 
The equatorial streamer is disrupted at this point. 
A fraction of the magnetic flux will remain in the equatorial region, forming a smaller streamer (Figure \ref{fig:field}, second panel, $t_2$). 
The rest of the magnetic flux reconnects and forms new plume/pseudo-streamer structures at mid-latitudes (as in \citealt{pinto_coronal_2010} and \citealt{wang_solar_2007}).
The letter B indicates the pseudo-streamer position.
New coronal holes now appear at low latitudes.
The polewards progression continues at a steady pace opening up its way by reconnecting with the open magnetic flux.
At about $t=3.5\un{yr}$ one of the magnetic arcades of each of the newly formed pseudo-streamer breaks and quickly opens up (Figure \ref{fig:field}, third panel, $t_3$, letter C).
As a result, a new coronal hole with inverse magnetic polarity rapidly forms.
This coronal hole will grow wider and fill up all the polar regions as the polarity inversions proceeds.
The previous coronal hole's field then closes down (Figure \ref{fig:field}, fourth panel, $t_4$),  and ends up disappearing below the surface 
(Figure \ref{fig:field}, fifth panel, $t_5$).
The corresponding coronal arcades (the ones closing down near the poles) are indicated by the letter D in the figure.
At $t = 11\un{yr}$ all traces of the previous coronal hole magnetic field have disappeared (Figure \ref{fig:field}, last panel, $t_6$).
The system is back to a state very close to its original state, but with the polarity of the magnetic field reversed (the dynamo model used here produces very regular and symmetric cycles).

Note that the polarity reversal happens \emph{quickly} in the corona, even if the underlying $\mathbf{B}^0$-field evolves \emph{slowly}.
The wind flow is responsible for the quick ``opening-up'' of field lines as the magnetic flux concentrations evolve slowly at the surface, and ultimately for the change of connectivity between contiguous regions. 
Furthermore, at the polar axis, the magnetic field's inversion occurs at different times at different heights. 
To better describe this property, the top panel in Figure \ref{fig:brpole} shows radial cuts of the polar magnetic field at the same moments as those shown in Figure \ref{fig:field}.
The bottom panel shows the temporal evolution of the polar field at different heights.
The $B_r$ sign-switch happens first at higher altitudes and proceeds downwards.
The downwards progression of the reversal of $B_r$ is quick ($\delta t$ of order of a few days) between $r=15\un{\rsun}$ and $r=2\un{\rsun}$, but from there on it slows down considerably.
The delay is $\delta t \approx 6\un{months}$ between $r=2\un{\rsun}$ and $r=1.6\un{\rsun}$, and about $1\un{yr}$ between $r=1.6\un{\rsun}$ and $r=1.3\un{\rsun}$.
The overall delay (between top and bottom) is about $4\un{yr}$.
This delay in the lower coronal layers is due to the slow disappearance of the coronal arcades near the poles, as shown in the last three panels of Figure \ref{fig:field}.

Figure \ref{fig:brlatitude} shows $B_r$ as a function of $\theta$ at different altitudes and at different moments of the cycle.
Faraway from the surface, $\mathbf{B}\approx B_r \mathbf{e}_r$ and is nearly independent of the latitude, except in the vicinity of a current sheet.
Close to the surface (beneath $r\approx 2-3\un{\rsun}$), the magnetic field organises in much more complex ways.
The potential component of the magnetic field dominates in the lower layers of the corona whereas the induced component largely dominates above (as the wind flow becomes stronger and approaches its asymptotic velocity).
From there on, the solar wind flow and the total magnetic field essentially align with the radial direction.

The multiple current sheets shown in the intermediate panels in Figure \ref{fig:field} may be interpreted as a highly warped current sheet in the real non-axisymmetric corona.

At the activity minimum the open flux is restricted to large polar coronal holes (about $30^\circ$ in latitude in each hemisphere).
This open flux will eventually fill all the available space at greater heights (spanning $180^\circ$ in latitude, from pole to pole), as the streamer thins out; above $r\approx 3\un{\rsun}$ all the magnetic field is open.
Closer to the maximum, the open flux sources are more spread in latitude but cover a smaller latitudinal extent altogether (about $5^\circ$ in latitude in each hemisphere).
These multiple thin coronal holes will nevertheless grow with height, and will end up filling up all available space. Above $r\approx 2.5\un{\rsun}$ all the magnetic field is open.
In other words, the average flux-tube expansion factors will be much higher at the maximum than at the minimum.

\section{Solar wind speed}
\label{sec:windspeed}

%%% fig 6
\begin{figure}[!ht]
  \centering
 \includegraphics[width=.95\linewidth]{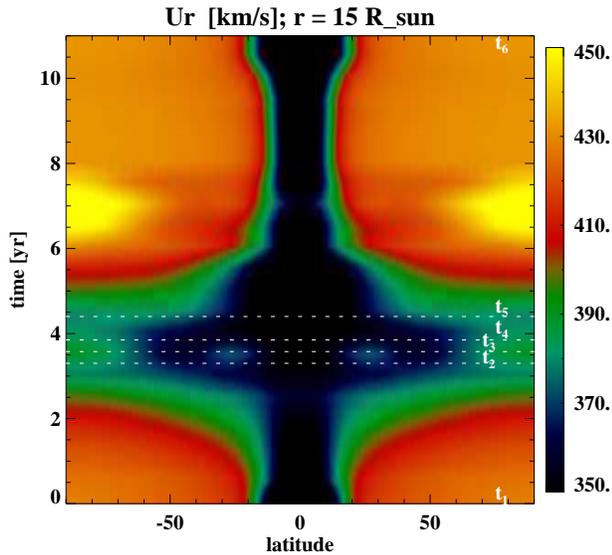}
    \caption{Solar wind speed at the outer boundary during the cycle as a function  of time and latitude.
      Most of the time, faster wind outflows occupy a large latitudinal extent ranging from the poles to latitudes as low as $\pm 20^\circ$.
      The exception is the polarity reversal phase ($t\in \left[2,5\right]\un{yr}$).
      Two low-latitude wind channels appear at about $t=3\un{yr}$, corresponding to the newly formed coronal holes (see Figure \ref{fig:field}, at $t_3$).}
  \label{fig:wind}
\end{figure}

We focus now on the variation of the solar wind velocity during the solar cycle, and on how it is distributed in latitude.
The evolution of the coronal magnetic topology (\S\ref{sec:magfield}) has a direct influence on the size and distribution of wind sources at the surface of sun, via changes in position and width.
The local magnetic pressure dominates over the wind's dynamical pressure at the lowest layers, and the local $\mathbf{B}$-field's amplitude and inclination mostly determine whether a given field line is \emph{open} or \emph{closed}.
Conversely, the latitudinal distribution of the solar wind faraway from the sun cannot be trivially predicted from the magnetic field's configuration at the surface. The solar wind flows accelerates along open field lines. A fluid element of cross-section $A_0$ at the surface will accelerate along a magnetic flux-tube with cross-section $A\left(r\right)$.
The final velocity profile depends on the flux-tube's expansion factor $\frac{A\left(r\right)}{A_0}$.
The expansion factor itself results from the competition between the wind's dynamical pressure and the magnetic pressure in the corona. 

The terminal wind speed and flux tube expansion factor are inversely correlated at all latitudes and at all times, agreeing with the well established 
Wang--Sheeley--Arge semi-empirical relation \citep{wang_solar_1990,arge_improvement_2000}.
In the initial acceleration phase (below $r\sim 3\un{\rsun}$) the expansion factors vary considerably, both in latitude and radius. At larger radii, though, the multiple and initially independent wind flows merge into a bulk spherical outflow (${A\left(r\right)}/{A_0}\propto r^{-2}$, independently of the latitude). Latitudinal inhomogeneities still subsist in this flow, being a result of the evolution (acceleration) of each parcel of wind flow along its path starting from the sun's surface.

Figure \ref{fig:wind} shows the latitudinal distribution of the solar wind speed
at $r=15\un{\rsun}$ (the domain's outer boundary) during the solar cycle.
The wind velocity vs. latitude diagram shows good qualitative agreement with 
the predictions made by \citet{wang_sources_2006} matching ULYSSES \emph{in situ} measurements and the more recent multi-station IPS (interplanetary scintillation) coronal observations by \citet{tokumaru_solar_2010}.
The separation between fast and slow wind is well visible in the figure.
At the activity minima, fast solar wind originates essentially from high latitude regions, while the slow wind flows mostly closer to the \emph{streamer} frontiers, at lower latitudes.
As the activity cycle progresses from the minimum to the maximum, the slow wind expands over towards the poles and takes over most of the latitudinal domain.
On the declining phase of the activity cycle, the fast wind recovers the polar regions and progressively extends towards lower latitudes, restraining the slow wind flow to the equatorial region.
Some irregularities appear above this otherwise too simple scenario.
Most remarkably, two short-lived wind channels appear at low latitudes between $t=3\un{yr}$ and $t=4\un{yr}$, that is, during the polarity inversion.
This can be seen in Figure \ref{fig:field} (panels $t_2,\ t_2,\ t_3$);
note how the blue shaded wind channel appears and evolves.
These flows originate in the newly formed mid-latitude coronal holes.
They follow the corresponding magnetic flux-tubes, bending over from $\sim 50^\circ$ at the surface down to $\sim 30^\circ$ at the outer boundary.
Although they seem to fade away quickly in Figure \ref{fig:wind}, they actually last till the declining phase of the activity cycle, but the wind flowing within these coronal holes slows down.
This slowing down is due to the increasingly higher expansion factor for the coronal hole, as can be seen in Figure \ref{fig:field}.
Note how the blue shaded coronal hole expands to fill the whole hemisphere faraway from the sun while its latitudinal extent at the surface remains approximately constant (from $t_3$ to $t_5$).
This continues while the polar closed-field regions progressively disappears; all open flux will merge into a wide polar coronal hole afterwards ($t_6$).
Also, the closing down of magnetic flux near the poles (during the cycle's decay phase, at about $t=7\un{yr}$) is related to the appearance of fast wind flows close to the polar axis.
These correspond to newly formed thin polar flux-tubes expanding almost radially, which will also merge afterwards into the wide polar coronal hole.

The wind's velocity values presented here are expected to be lower than the values measured \emph{in situ} near the Earth's orbit. 
The reader should note that, on one hand, our numerical domain extends only up to $15\un{\rsun}$, and that the wind flow has not yet reached its asymptotic velocity at this height.
Nevertheless, the relative variations of $V_r$ in latitude should not change considerably.
The wind undergoes a purely spherical expansion between $15\un{\rsun}$ and $1\un{AU}$.
On the other hand, this does not fully accounts for the low velocities values found, though.
We could expect velocities in the order of $500-600\un{km/s}$ from this model if the domain extended up to $1\un{AU}$.
Faster wind velocities require a complete and consistent energetic treatment, which is beyond the scope of the current paper.
Besides, \citet{pinto_time-dependent_2009} show that the anti-correlation between terminal wind velocity and flux-tube expansion factor is still verified in self-consistent non-isothermal cases. 
The only strong constraint over the domain's radial extent regarding the physical correctness of the numerical model is that it has to completely contain all critical surfaces (sonic and  alfv\'enic; \emph{cf.} \S\ref{sec:dip}), which it always does.

\section{Mass flux, momentum flux and magnetic braking torque}
\label{sec:massflux}

%% fig 7
\begin{figure}[h]
  \centering
  \includegraphics[width=\linewidth]{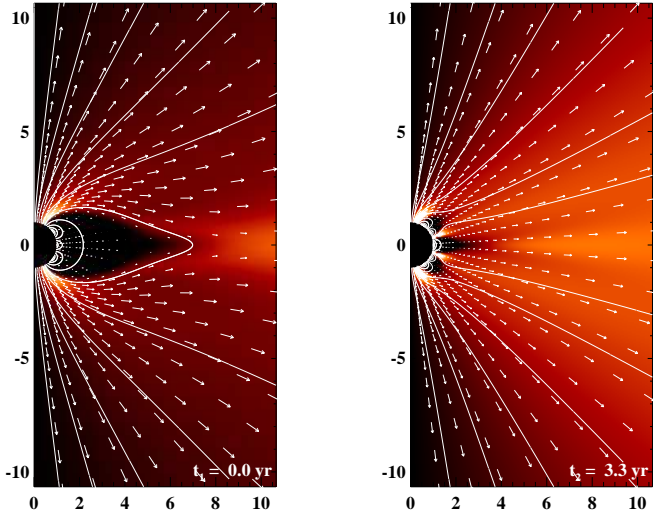}
  \caption{Mass flux $\rho V_r r^2 \sin\theta$ in the
    meridional plane at $t=0\un{yr}$ (left) and
    at $t=3.3\un{yr}$, about the polarity inversion (right), that is, respectively, instants $t_1$ and $t_2$.
    The factor $r^2\sin\theta$ 
    is due to the spherical expansion of a surface element
    normal to the radial direction. Outflows which
    originate at lower latitude dominate the global mass loss
    rate.}
  \label{fig:massflux}
\end{figure}

%% fig 8
\begin{figure}[h]
  \centering
  \includegraphics[width=.95\linewidth]{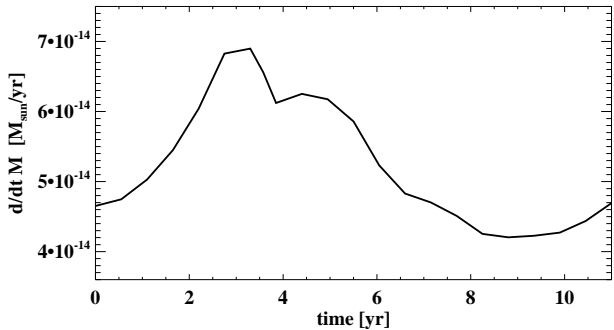}
  \caption{Mass loss rate $\dot{M}$ during the solar cycle.
  }
  \label{fig:dotm}
\end{figure}

%% fig 9
\begin{figure}[h]
  \centering
    \includegraphics[width=\linewidth]{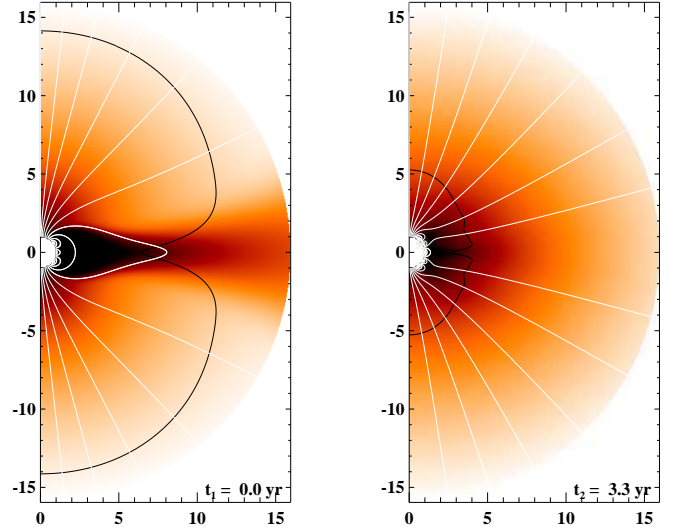}
  \caption{Alfv\'en surfaces (black contours) at the same instants as in
    fig. \ref{fig:massflux} (instants $t_1$ and $t_2$).
    White lines are magnetic field lines, and the
    colorscale represents the wind's poloidal Mach number.}
  \label{fig:alfvensurf}
\end{figure}

%% fig 10
\begin{figure}[h]
  \centering
  \includegraphics[width=.95\linewidth]{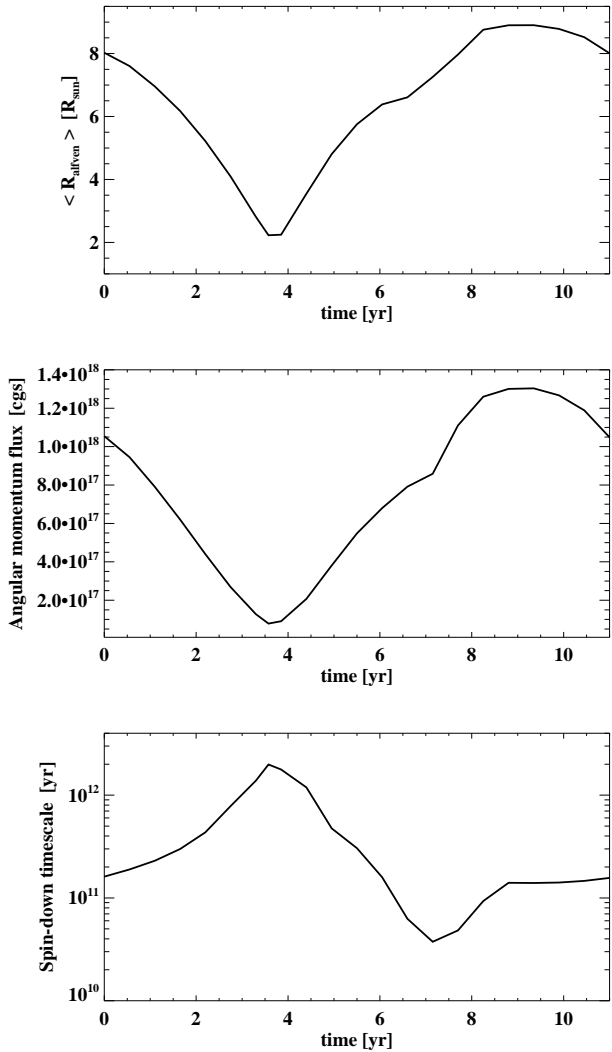}
  \caption{Mean Alfv\'en radius $\langle r_A\rangle$ (top), specific angular momentum flux $\Omega \langle r_A^2\rangle$ (middle) and magnetic braking time-scale $\delta t_{sd} = {J_\odot}/{\dot J}$ (bottom) during the solar cycle. $J_\odot$ is the sun's angular momentum \citep{gilman_angular_1989}; $\dot J$ is the wind's angular momentum loss rate.}
  \label{fig:ralfven}
\end{figure}

Figure \ref{fig:massflux} shows the radial mass flux associated with the solar wind in the corona at two different instants of the solar cycle.
In both cases, the net mass outflow in the polar coronal holes is higher near the streamer boundaries than closer to the poles.
At the maximum of activity, thin coronal holes appear also at low latitudes (as described previously).
Despite their small latitudinal extent at the surface, the associated mass flux is important when compared to that in the polar coronal holes.
For an outflow with a given latitudinal extent $\delta\theta$, the actual surface area it crosses is equal to $2\pi r^2 \sin\theta \delta\theta$.
Therefore, low latitude coronal hole are more prone to produce higher mass outflow rates.
Arguably, this is a consequence of the axi-symmetrical nature of our model.
Nevertheless, the real three-dimensional sun shows some degree of axi-symmetry in the distribution of the streamers and coronal holes.
These low-latitude wind streams correspond to the two channels visible in Figure \ref{fig:wind}, at about $t=3\un{yr}$.

Figure \ref{fig:dotm} shows the total mass loss rate
\begin{equation}
  \dot{M} = 2\pi R_0^2 \int_0^\pi \rho V_r
  \sin{\theta} d\theta\
  \label{eq:dotm}
\end{equation}
evaluated at the outer boundary of the numerical domain.
The mass loss rate evolves in par with the activity cycle.
That is, $\dot{M}$ is maximal at about $t=3\un{yr}$ (during the activity maximum) and minimal during the activity minima.
The amplitude of the mass loss rate varies by a factor of about $1.6$ in time, from $4.2\e{-14}\un{M_{\odot}/yr}$ at the activity minimum and $6.9\e{-14}\un{M_{\odot}/yr}$ at the activity maximum.
This trend supports the idea that the lower latitude outflows --- which appear mainly close to the activity maximum --- contribute with higher net mass outflow rates, as discussed in the previous paragraph.
Note that the increase in $\dot{M}$ cannot be due to the variations in the velocity of the wind, as it is actually lower at almost all latitudes during the activity maxima (see Figure \ref{fig:wind}) and contributes to lowering the net mass flux.
On the other hand, the variations of $\rho$ at the lower numerical boundary are small throughout all the cycle.
  The amplitude of the density fluctuations at the surface within open-field regions is always below $6\%$ of its average value (amounting to at most $10\%$ if both open and closed-field regions are considered altogether).

The way the open flux maps from the solar surface up the outer domain must therefore be the main cause for the variations found in the mass loss rate.

The solar wind outflow carries angular momentum away from the sun.
The specific angular momentum flux, magnetic braking torque and spin-down time-scale are deduced from the Alfv\'en surface's geometry and solar rotation rate $\Omega_0$.

The Alfv\'en surface is the geometric locus where the wind velocity equals the Alfv\'en speed $c_A = {B}/{\sqrt{4\pi\rho}}$.
The Alfv\'en radius $r_A$ is the cylindrical distance from the rotation axis to this surface.
Classical wind theory \citep[][and many others]{weber_angular_1967} states that the angular momentum balance problem can be simplified as follows. 
The plasma inside the Alfv\'en surface is kept in solid rotation while flowing outwards along magnetic field lines. 
The plasma flow then becomes super-alfv\'enic and its angular momentum is conserved thereafter. 
The physical process responsible for maintaining the solid rotation while $V < c_A$ is the magnetic tension, which works against the tendency for the field-lines to spiral backwards in the azimuthal direction (keeping them as straight as possible). Beyond the critical surface, the magnetic tension loses its efficiency.

Of course, in the real sun the transition from rigid to non-rigid rotation is a continuous and smooth one.
The azimuthal magnetic field does not suddenly change from straight to spiralled.
But quantitatively, it all works out \emph{as if} there was a sharp transition.
Following this picture, one can define an average Alfv\'en radius $\langle r_A \rangle$ --- an ``effective lever arm length''. 
The specific angular momentum flux rate is then
\begin{equation}
  l = \Omega_0 \langle r_A^2 \rangle\ .
  \label{eq:mompermass}
\end{equation}
The resulting torque applied on the sun is \citep{matt_accretion-powered_2008-1}
\begin{equation}
  \tau = -\dot{M} \Omega_0 \langle r_A^2 \rangle\ .
  \label{eq:tau}
\end{equation}

The angular momentum per unit volume $J$ of a parcel of solar wind plasma rotating with azimuthal velocity $v_\phi$ is
\begin{equation}
  J_w = \rho r \sin{\theta} v_\phi\ .
  \label{eq:jw}
\end{equation}

The angular momentum per unit volume crossing a surface element $dA$ is then $\dot{J}_w = J_w v_r dA$. Integrating over a spherical surface of radius $r_0$ and assuming axi-symmetry translates into
\begin{equation}
  \dot{J}_w = 2\pi r_0^3 \int_0^\pi \rho  v_r v_\phi \sin^2\theta d\theta\ .
  \label{eq:jdot}
\end{equation}
We then define the magnetic spin-down time-scale as
\begin{equation}
  \delta t_{sd} = \frac{J_\odot}{\dot{J}_w}\ ,
  \label{eq:dtsd}
\end{equation}
where $J_\odot$ is the sun's angular momentum. We estimated
\begin{equation}
  J_\odot = \frac{8\pi}{3}\Omega_0\int_0^{R_{\odot}} \rho\left(r\right) r^4 dr \approx 1.84\e{48}\un{g\ cm^2\ s^{-1}}
  \label{eq:jsun}
\end{equation}
\citep{gilman_angular_1989,stix_sun:_2002} using a seismically calibrated solar model for $\rho\left(r\right)$ \citep[a CESAM model,][]{morel_cesam:_1997,brun_seismic_2002}.

The main difficulty now lies in the definition of $\langle r_A \rangle$.
Figure \ref{fig:alfvensurf} shows the Alfv\'en surface at two different instants of the cycle ($t=0,3\un{yr}$).
This critical surface shows a regular shape for most of the activity cycle, being close to spherical at most latitudes (especially, higher latitudes).
At low latitudes, though, the Alfv\'en surface approaches a more cylindrical shape.
Some irregularities appear as inward incursions as $\mathbf{B}$ vanishes in current sheets (there's a small but finite outflow, so ${v}/{c_A}\rightarrow\infty$ there).
The most evident example of such an excursion corresponds to the equatorial streamer (\emph{e.g} Figure \ref{fig:alfvensurf}, left panel).
We defined here the average Alfv\'en radius $\langle r_A\rangle$ as the average cylindrical radii of the Alfv\'en surface (that is, $r_A$, the distance to the axis) weighted by the local mass flux $r^2\sin\theta\rho\mathbf{v}$ crossing the surface
\begin{equation}
  \label{eq:ralfven}
  \langle r_A \rangle = 
  \frac{\int r^2\sin\theta \rho\mathbf{v}\cdot 
    r_A\left(\theta\right)\hat{\mathbf{n}} d\theta}
  {\int r^2\sin\theta \| \rho\mathbf{v}\|d\theta}\ .
\end{equation}
The sections of the Alfv\'en surface crossing heliospheric current sheets are discarded from the computation (that is, the inwards incursions described above).
The average Alfv\'en radius $\langle r_A\rangle$ was found to be
about $9\un{R_{\odot}}$ at the minimum and $2.2\un{R_{\odot}}$ at the maximum (the time average being about $6.3\un{R_{\odot}}$).
Figure \ref{fig:ralfven} (top panel) shows the average Alfv\'en radius as  a function of time during the whole solar cycle.
The Alfv\'en radius correlates well with the global magnetic field's amplitude for the most part of the cycle, but shows  a negative correlation with the polar surface magnetic field (compare with Figure \ref{fig:brpole})

The values found for $\langle r_A \rangle$ were then used to compute the solar spin-down time-scale (eq. \ref{eq:dtsd}, with $r_0 = \langle r_A \rangle$ in eq. \ref{eq:jdot}) and the specific angular momentum flux (eq. \ref{eq:mompermass}, using $\langle r_A^2 \rangle$ rather than $\langle r_A\rangle$).
The results are displayed in Figure \ref{fig:ralfven} (middle and bottom panels) as a function of time. 
The spin-down time-scale $\delta t_{sd}$ due to the breaking torque exerted by the solar wind was found to vary between $3.7\times 10^{10}\un{yr}$ and $2\times 10^{12}\un{yr}$ (the time average being $4.7\times 10^{11}\un{yr}$).
The quantities $\langle r_A\rangle$ and $\delta t_{sd}$ show a strong negative correlation, meaning the ``effective lever arm length'' mostly determines the breaking efficiency.
Note that the mass flux rate evolution throughout the cycle works against this trend (\emph{cf.} Figure \ref{fig:dotm}): the highest mass outflow rate occurs at about the activity maximum, but its effect is overwhelmed by the large variations of $\langle r_A\rangle$ during the activity cycle.

\section{Discussion}
\label{sec:disc}

We have shown that the coronal magnetic field's topological variations due to the solar dynamo have a strong influence over the solar wind's velocity and associated mass and momentum fluxes.
The ratio of open to closed magnetic flux at the sun's surface also varies. 
These elements altogether define the variations of the Alfv\'en radii (both in latitude and in time) and how effectively the solar wind carries angular momentum away from the sun.

The average Alfv\'en radius $\langle r_A\rangle$ was found to be about $9\un{R_{\odot}}$ at the activity minima and $2.2\un{R_{\odot}}$ at the maximum.
The global poloidal magnetic field's amplitudes follow the same trend.
Unexpectedly, though, the polar magnetic field amplitudes are anti-correlated with the global field's amplitudes.
The polar magnetic field (at the surface) happens to be stronger at the moments when the closed to open flux ratio is higher.
As a consequence, the polar magnetic flux-tubes show higher expansion rates at these moments of the cycle (higher expansion factors imply stronger $\mathbf{B}$-field decays).
This behaviour could be related to the nature of the particular dynamo model used in this work, which is a simplified one (as exposed in \S\ref{sec:stelem}).
We should nevertheless note that, in spite of its simplicity, the model reproduces well many solar coronal features.
Namely, the topological characteristics of the poloidal magnetic field we obtained are consistent with those found by \citet{wang_topological_2003} (using a different type of model for the surface field).
Also, the latitudinal distribution of the wind velocities during the cycle we computed agree with those proposed by \citet{wang_sources_2006} (well matched by \emph{ULYSSES} data; see their Figure 3) and those by \citet{tokumaru_solar_2010} (built from IPS data collected during $2$ solar cycles).

The mass loss rate (as carried away by the wind) we computed averages out in time to $\sim 5.3\e{-14}\un{M_{\odot}/yr}$, 
a value comparable to in-situ measures near the Earth (giving a standard value of about $10^{-14}\un{M_{\odot}/yr}$ if the wind flow was spherically symmetric).
Conversely, the solar spin-down time-scale due solely to the torque exerted by the magnetised wind outflow averages out to value of the order of $10^{11}\un{yr}$, above the standard $10^{10}\un{yr}$.
%%%%

One reason for this disparity may lie in the fact that we used a one-fluid isothermal MHD wind model.
Considering a proper (i.e, non-isothermal) treatment of the energetics of the wind could lead to different quantitative estimations on the wind velocities.
For example, \citet{pinto_time-dependent_2009} found a broad range of wind velocities in their configurations including a chromosphere -- corona transition region and a coronal heating flux dependent on the local magnetic field's strength. 
The actual asymptotic wind velocities depended strongly on the radial distribution of the mechanical heat flux dissipation.
Under the right conditions, a non-isothermal two-fluid approach can also produce higher bulk velocities \citep[see][]{endeve_coronal_2001,grappin_two-temperature_2011}.
An additional word of caution should be given in respect to the choice of a constant and uniform coronal temperature.
It is known that the mass flux in an isothermal wind flowing within a given magnetic flux-tube depends on the coronal temperature $T$ as
\begin{equation}
  \label{eq:parkermassflux}
  \rho v \propto \rho_0 T^{-3/2} \exp\left(-C/T \right) \times f\left(r\right)\ ,
\end{equation}
\citep[cf.][among others]%
{parker_dynamical_1964-2,leer_constraints_1979,hansteen_coronal_1995}
where $\rho_0$ is the density at the base of the flux-tube, $C$ a constant and $f\left(r\right)\propto A\left(r\right)/A_0$ a geometrical factor describing the radial expansion of the flux-tube.
The absolute values found for the wind's mass flux are therefore very sensitive to the choice of the parameter $T$, while the amplitude of its variations during the cycle (keeping $T$ constant) are only due to the geometrical factor $f\left(r\right)$ and the (small) variations in $\rho_0$.
Additionally, it has been suggested that in non-isothermal scenarios the mass flux scales linearly with the mechanical heat flux injected at the surface \citep{leer_constraints_1979,hammer_energy_1982,withbroe_temperature_1988,hansteen_coronal_1995}.
We emphasize that, nevertheless, the results described in this manuscript depend essentially on other parameters of the problem, namely the cyclic variations in the global coronal topology and the consequent open flux-tube expansion factors (cf. the function $f\left(r\right)$ above).
The isothermal approximation should in this case provide a good assessment of the variability of the global wind properties during the activity cycle.
Future work will address these issues by taking into account a more complete (i.e, non-isothermal) treatment of the energetics of the wind.

%%%%
Also, we neglected any effects due to the rotation on the wind's velocity (which should be small, anyway, as the sun is a slow rotator).

The most remarkable feature found is the large temporal variability of these quantities (Alfv\'en radius and spin-down time-scale) over an activity cycle.
The sun loses mass and momentum non steadily, which translates into variable braking torques applied to the surface.
We should add that the latitude and the width of the open-field and wind sources at the surface also vary in time, meaning the torque is applied at different places at different times of the cycle.
During most of the activity cycle, the open-field crosses the solar surface at high latitudes only.
The exception is the interval about the polarity inversion, when the polar coronal holes are greatly reduced and low latitude open-field flux-tubes appear.
We will consider whether this effect can be translated into meaningful upper boundary conditions (forcing, mass and momentum sources/sinks) for the solar convection and dynamo processes in future research.

%% fig 11
\begin{figure}[!h]
  \centering
  \includegraphics[width=0.95\linewidth]{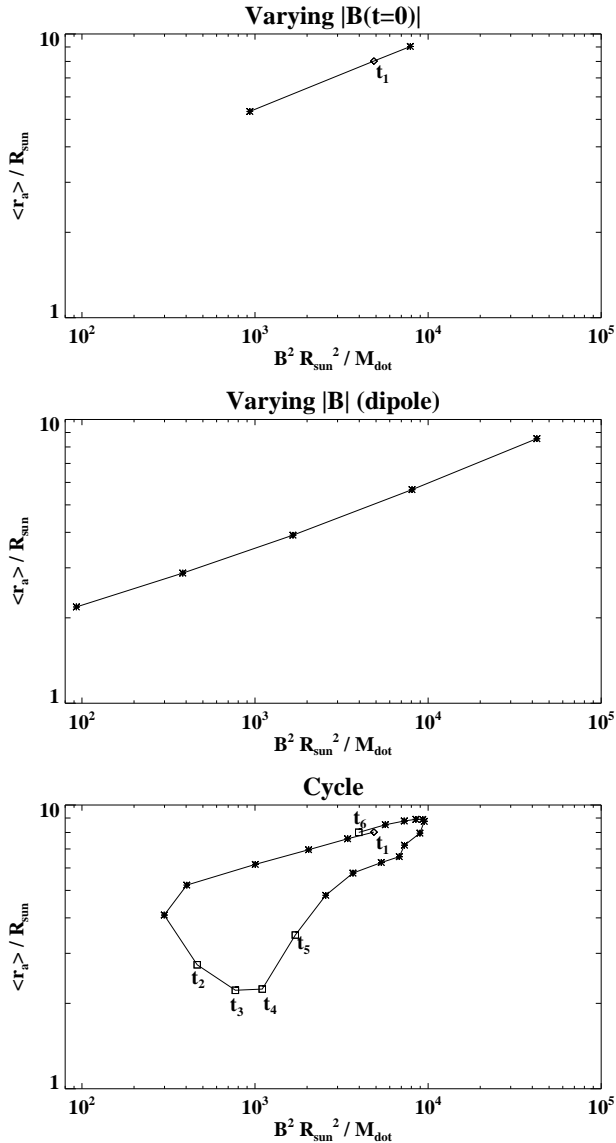}
  \caption{Mean Alfv\'en radius as a function of $\Upsilon$.
    Top: variation of the amplitude of the external magnetic field for our $t_1=0\un{yr}$ case.
  Middle: dipolar benchmark case (varations of the external field's amplitude, as above).
  Bottom: the same relation, but throughout the cycle.
  The labels $t_1$ to $t_6$ identify the instants shown in Figure \ref{fig:field}.
  The label $t_1$ appears both on the top and bottom panels, as it corresponds to the same state.
}
  \label{fig:upsilon}
\end{figure}

Now, as the solar cycle progresses the effects of an ever-changing magnetic field topology add up to the variations of the global amplitude of the field.
This makes it hard to distinguish between their respective contributions.
Let us first clarify the status of our coronal model in regard to this issue.
We ran our model through $2$ additional test cases in which we varied only the external magnetic field's amplitude, and not its topology (i.e, the potential magnetic field only varied by a multiplicative factor between consecutive runs).
In the first case, we took the first element of our time-series ($t=0$ in the figures in the preceding sections) and varied the amplitude of the external field by a constant multiplicative factor.
The second case corresponds to a purely dipolar external field $\mathbf{B}^0$ whose amplitude we also varied, letting us perform a comparison with previous studies.
We then computed the mean Alfv\'en radius $\langle r_A \rangle$ and the mass loss rate $\dot{M}$ for each new run and tested against the correlation
\[\frac{\langle r_A \rangle}{R_\odot} \propto \Upsilon^m = \left[ \frac{\left(\|\mathbf{B}\| R_\odot\right)^2}{\dot{M}} \right]^m\]
suggested by \citet{matt_accretion-powered_2008-1}, where $\|\mathbf{B}\|$ is the global field's amplitude.
This scalar quantity is well defined for any given analytical external magnetic field but not for the ones obtained numerically from the dynamo model.
We chose here to associate it to the unsigned magnetic flux integrated over the solar surface (after comparing both definitions for the dipolar test cases).
Figure \ref{fig:upsilon} (top and middle panels) shows that the expression above fits our $2$ test cases for a power law index $m \approx 0.2$, which is in good agreement with what \citet{matt_accretion-powered_2008-1} found for their dipolar case.
This result validates the (rather large) variations in Alfv\'en radius we found.
In addition, we plotted the same relation for the computations running through the whole cycle (Figure \ref{fig:upsilon}, bottom panel).
The same power law is still observed for most of the cycle, but it gets slightly steeper just before the polarity inversion and then deviates away from it (flipping sign) immediately after (between $t=3$ and $t=5\un{yr}$), drawing a closed cycle in the diagram.
This result suggests that the Alfv\'en radius correlates positively with ${\|B\|^2}/{\dot{M}}$ when the dipolar component of the open magnetic field is strong (which happens during most of the cycle), but may correlate negatively for some other higher order geometries.

As expressed earlier in the text, we computed a succession of steady-state wind solutions rather than the detailed dynamical evolution of a particular event.
One justification for this choice is of numerical nature: the disparity of time-scales for the magneto-convective and coronal phenomena is very large.
The second and perhaps more important justification is that varying the potential magnetic field in time introduces a non causal perturbation to the system.
That would correspond to an instantaneous propagation of a perturbation to the field sources.
This violation of causality remains even for a continuous and arbitrarily slow evolution of the background potential field.
Note that localised perturbations to a MHD system are propagated away with phase velocities which correspond to the MHD wave modes. 
These can be arbitrarily small (even null in some places) and very anisotropic.
To work around this issue one would need to include the field sources in the domain (and then self-consistently compute the system's response to their perturbation).
Alternatively, one could emulate the proper physical behaviour of such a system by injecting/propagating the hydro-magnetic perturbations through the numerical boundaries.
This issue will be the subject of future work.

\section{Summary and conclusions}
\label{sec:summary}

We have performed MHD numerical simulations coupling a solar dynamo model with a corona and solar wind model throughout a complete activity cycle.
In short, we found that:
\begin{itemize}
\item The latitudinal distribution of the asymptotic wind velocities is sensitive to the magnetic topology as it varies during the solar cycle. The fast wind - slow wind pattern shows good qualitative agreement with those in 
\citet{wang_sources_2006} and
%\citet{manoharan_solar_2010} and
\citet{tokumaru_solar_2010},
as shown in Figure \ref{fig:wind}.
\item  The polarity reversal happens rather abruptly in the corona, in contrast with the progressive evolution of the solar wind's velocities and of the surface magnetic field (Figure \ref{fig:brpole}).
\item Sun's global mass loss rate, Alfv\'en radius and  momentum flux
  all vary considerably throughout the cycle (Figures \ref{fig:dotm} and \ref{fig:ralfven}).
  The dominant causes are the position and latitudinal extent of the photospheric sources of solar wind and the geometry of the Alfv\'en surface.
\item The zones of application of the braking torque due to the wind vary in time.
  Overall, the wind's breaking torque should contribute to slow down the surface layers at high latitudes, but regions of application of torque appear occasionally at lower latitudes.
\end{itemize}

Future work will focus on testing other types of solar dynamo models, using Babcock-Leighton flux transport based on large scale meridional circulation 
\citep{dikpati_diagnostics_2004,jouve_role_2007}
or turbulent magnetic pumping
\citep{guerrero_turbulent_2008}, 
as well as non-isothermal winds.
We will furthermore consider the formation of a non-rigid coronal rotation and its effects on the solar wind properties.

\acknowledgements
  This work was supported by the ERC Grant \#207430 (STARS2 project, PI: S. Brun, \\ http://www.stars2.eu) and the CNRS PNST Interfaces group. %
  Computations were carried out using CNRS IDRIS and CEA's CCRT facilities.
  We thank S. Matt for useful discussions.

%\clearpage

% \bibliographystyle{apj}
% \bibliography{refs}

\end{document}